\documentclass[doublecol]{epl2}
\usepackage{amsmath}
\usepackage{amsfonts}
\usepackage{amssymb}
\usepackage{graphicx}
\usepackage{supertabular}
\usepackage{xcolor}

\newcommand{\aee}{{\rm Tr}A}
\newcommand{\be}{{\rm Tr}A^{2}}
\newcommand{\ee}{{\rm Tr}A^{3}}
\newcommand{\de}{{\rm Det}A}

\title{Quadratic superconducting cosmic strings revisited\footnote{Contribution
under the title of ``\textit{Exact 5-Dimensional Cosmic String
Solutions}" to the International Conference on Dynamics and
Thermodynamics of Black Holes and Naked Singularities II, Milan,
Italy, May 2007.}} \shorttitle{Quadratic superconducting cosmic
strings revisited}
\author{Mustapha Azreg-A\"{\i}nou\footnote{azreg@baskent.edu.tr}}
\shortauthor{M. Azreg-A\"{\i}nou}

\institute{
   Ba\c{s}kent University, Engineering Faculty, Ba\u{g}l\i
ca Campus, Ankara, Turkey}


\abstract{It has been shown that 5-dimensional general relativity
action extended by appropriate quadratic terms admits a singular
superconducting cosmic string solution. We search for cosmic
strings endowed with similar and extended physical properties by
directly integrating the non-linear matrix field equations thus
avoiding the perturbative approach by which we constructed the
above-mentioned \textsl{exact} solution. The most general
superconducting cosmic string, subject to some constraints, will
be derived and shown to be mathematically \textsl{unique} up to
linear coordinate transformations mixing its Killing vectors. The
most general solution, however, is not globally equivalent to the
old one due to the existence of Killing vectors with closed
orbits.}

\begin{document}

\maketitle

\section{Introduction}
In \textit{d}-dimensional general relativity (GR) or Kaluza-Klein
(KK) theories ($d>4$), one can generate non-linear electrodynamics
just by coupling the usual scalar curvature to quadratic and
higher order terms in the curvature making up the Gauss-Bonnet
(GB) term. It has been shown that for (d $>4$) the inclusion of
these extra terms in the Lagrangian leads to modified, most
general field equations including up to second-order derivatives
of the metric \cite{Lovelock}, which upon 4+($d-$4) dimensional
reduction split into a set of equations governing non-linear
4-dimensional GR coupled to both non-linear electrodynamics and
scalar fields \cite{pio,Zumino,Madore85,MH,DM86,Wheeler,Azreg}.
The effective reduced equations are not equivalent to those
obtained upon coupling 4-dimensional GR to non-linear
electrodynamics \cite{Ayon1,Ayon2,Burinskii}, nor are they
equivalent to those derived from higher dimensional
Einstein-Maxwell or Einstein-Born-Infeld theories modified by GB
term \cite{Buch,Wiltshire,Thi,Cai}.

Some solutions to these extended \`{a} la GB KK theories under the
spherical and cylindrical symmetry assumptions were constructed
and interpreted as singular static black holes~\cite{Wheeler} and
singular 4-stationary neutral, charged or superconducting cosmic
strings as well as multiple cosmic strings~\cite{Azreg},
respectively. Regular wormhole-type solutions have been
constructed too \cite{Kar,024002,Thi}, some of which do not
violate the weak energy condition \cite{Kar} and others do not
violate any energy condition \cite{024002}.

Generalized \`{a} la GB GR theories have gained their sharp appeal
these last three years with ever-decreasing efforts in dealing
with up-to-date problems of black holes, brane world models,
gravitational dark energy, cosmic acceleration and string theory
\cite{024002,Thi,024018,044007,084025,024042,086002,104003,123515,Berej,Cai}.
This parallels a renewed interest in string theory fueled by the
discovery of Capodimonte-Sternberg-Lens Candidate, CSL-1, a pair
of aligned galaxies which lie at a redshift of z = 0.46 whose
double image could be a result of gravitational lensing caused by
a cosmic string~~\cite{Sake}. The connection is that the GB
Lagrangian (GBL) appears in the low energy limit of string theory
and its linearized form (GBL) leads upon quantization to
ghost-free theory \cite{pio,086002}.

In the second section the KK theory extended by GB term is
introduced and the field equations under the cylindrical symmetry
assumption are derived. The cosmic-string solutions obtained so
far are discussed and classified. In the third section we derive
by a cognitive approach the most general superconducting
cosmic-string solution subject to some constraints and prove its
unicity up to similarity transformations. In the fourth section we
discuss the physical content of the new solution and conclude in
the last section.

\section{Quadratic Kaluza-Klein theory}
In 5~-~dimensional GR the most general field equations containing
at most second order partial derivatives of the metric write upon
ignoring a cosmological term
\begin{equation}\label{field}
    R_{AB} - (1/2)Rg_{AB} + \gamma L_{AB} = 0\,,
\end{equation}
where $\gamma$ is a constant and $L_{AB}$ is the covariantly
conserved ($L^{AB}{}_{;A}=0$) Lanczos tensor associated with the
GBL density ${\cal L}$

\begin{eqnarray}
  L_{AB} &\equiv& R_{A}{^{CDE}}R_{BCDE} - 2R^{CD}R_{ACBD} \nonumber \\
   & & -\,
2R_{AC}R_{B}{^{C}} + RR_{AB} - (1/4)g_{AB}{\cal L}\,,
\end{eqnarray}
with ${\cal L}\equiv R^{ABCD}R_{ABCD}-4R^{AB}R_{AB}+R^{2}$
($R_{ABCD}$ \& $R_{AB}$ are the 5-dimensional Riemann \& Ricci
tensors and $A,B,C,D,E=1,\ldots,5$).

From now on we restrict ourselves to stationary cylindrically
symmetric 5-dimensional metrics. Such metrics have four commuting
Killing vectors ${\xi}_a{}^A={\delta}_a{}^A$ ($a,b=2,\ldots,5$),
one of which (${\xi}_4{}^A$) is timelike, and two of which
(${\xi}_2{}^A$ \& ${\xi}_5{}^A$) have closed orbits. Using the
coordinates adapted to these vectors, the 5-metric can be
parameterized as
\begin{equation}\label{metric}
    {\rm d}s^{2} = -\,{\rm d}{\rho}^{2} + {\lambda}_{ab}(\rho)\,
    {\rm d}x^{a}\,{\rm d}x^{b}\,,
\end{equation}
where $x^2=\varphi$ \& $x^5$ are periodic, $x^4=t$ is timelike,
$x^3=z$ and $\rho$ is a radial coordinate. $\lambda_{ab}(\rho)$ is
a 4$\times$4 real symmetrical matrix of signature (-- -- + --) so
that the 5-metric has a Lorentzian signature (-- -- -- + --).
Introducing the matrices
$\chi\equiv{\lambda}^{-1}{\lambda}_{,\rho}$ \&
$B\equiv\chi_{,\rho}+(1/2)\chi^{2}$, eqs. (\ref{field}) reduce to
a system of non-linear scalar \& matrix differential equations
\begin{eqnarray}
\label{s} \mbox{\hspace{-11mm}}& &6\,{\rm Tr}B+({\rm
Tr}\chi)^{2}-{\rm Tr}\chi^{2}+\gamma \left\{{\rm
Tr}(B\chi^{2})-{\rm
Tr}(B\chi){\rm Tr}\chi \right. \nonumber \\
\mbox{\hspace{-11mm}}& &\left.+\,(1/2){\rm Tr}B[({\rm
Tr}\chi)^{2}-{\rm Tr}\chi^{2}]\right\}=0 \,,\\
\label{m} \mbox{\hspace{-11mm}}& &2\chi_{,\rho}+4{\rm
Tr}\chi_{,\rho}+({\rm Tr}\chi)\chi+ {\rm Tr}\chi^{2}+({\rm
Tr}\chi)^{2}\nonumber \\
\mbox{\hspace{-11mm}}& &+\,\gamma \left\{(\chi^{3})_{,\rho} -
({\rm Tr}\chi)(\chi^{2})_{,\rho}+[({\rm Tr}\chi)^{2}-{\rm
Tr}\chi^{2}]\chi_{,\rho}\right. \nonumber \\
\mbox{\hspace{-11mm}}& &-\,({\rm Tr}\chi_{,\rho})[\chi^{2}-({\rm
Tr}\chi)\chi]-(1/2)[({\rm Tr}\chi^{2})_{,\rho}\chi -
({\rm Tr}\chi)^{3}\chi]\nonumber \\
\mbox{\hspace{-11mm}}& &\left.+\,(1/2)[({\rm
Tr}\chi)\chi^{3}-({\rm Tr}\chi^{2})\chi^{2} -({\rm Tr}\chi)({\rm
Tr}\chi^{2})\chi]\right\}=0 \,.
\end{eqnarray}

The system (\ref{s} \& \ref{m}) remains invariant if one performs
a linear coordinate transformation with constant coefficients
mixing the four commuting Killing vectors together and their
associated cyclic coordinates
\begin{equation}\label{L}
    x^{a}=S^{a}{_{b}}x_{\rm N}^{b}\,,
\end{equation}
where $S^{a}{_{b}}$ is a constant real matrix. Here $x^{a}$ \&
$x_{\rm N}^{b}$ are the old and new coordinates, respectively.
Such a transformation is equivalent to a similarity transformation
on $\chi$ ($\chi=S\chi_{\rm N}S^{-1}$). Solutions related by such
transformations belong actually to the same class of equivalence.
However, when some Killing vectors have closed orbits, say
${\xi}_2{}^A$ and ${\xi}_5{}^A$ in our case, it is possible to
generate new solutions which are not globally equivalent to old
ones if one at least of these two vectors is re-scaled or mixed
with the others as a result of the transformation~\eqref{L}. For
instance, we have shown in sect. 4 of ref.~\cite{Azreg} how to
obtain a magnetic spacetime from Minkowski spacetime upon
performing a similarity transformation~\eqref{L} on the latter.

For the case $\gamma=0$, corresponding to pure KK theory, the
whole set of exact solutions to the system (\ref{s} \& \ref{m})
have been systematically constructed, classified and interpreted
as neutral or charged cosmic strings \cite{Azreg}. We found that
two classes~\eqref{a} among these solutions solve trivially
(\ref{s} \& \ref{m}) for $\gamma\neq0$ since their Lanczos term
vanishes identically; furthermore, they have been shown to be
unique mathematical solutions up to linear coordinate
transformations~\eqref{L}. A non-trivial solution with
non-vanishing lanczos term ($\gamma\neq0$) has been derived by a
perturbative approach, however, turned out to be exact and
interpreted as a superconducting cosmic string~\eqref{abd}. The
purpose of this note is to derive the most general superconducting
cosmic-string solution by directly integrating the non-linear
field equations (\ref{s} \& \ref{m}) under the same assumptions we
made in our previous work and prove its unicity from a
mathematical point of view, which has been left open so far.

For the case $\gamma\neq0$, we mainly focused on a) the search for
solutions depending on a scalar function $\omega(\rho)$ and a
constant real matrix $A$ of the form $\chi=\omega(\rho)A$. We have
shown that the mathematically unique solution is~\cite{Azreg}
\begin{eqnarray}
  \label{a} \mbox{\hspace{-9mm}}& & \chi = (2/\rho)A\,, \;
  \aee =\be =\ee =1\,,\;\de =0\,, \nonumber \\
  \mbox{\hspace{-9mm}}& &{\rm with}\;\;{\rm r}(A)=1\;\;{\rm or}\;\;
  {\rm r}(A)=2\,,
\end{eqnarray}
where r($A$) denotes the rank of $A$. The solution with rank~1 has
been interpreted as a neutral cosmic string and that with rank 2
as a charged cosmic string. For both strings the Lanczos term
vanishes identically so they are trivial solutions. Notice that
because of~\eqref{a} any similarity transformation on $A$, induced
by a coordinate transformation of the form~\eqref{L}, results in a
similarity transformation on $\chi$ and conversely. A result from
matrix theory confirms that two matrices related by a similarity
transformation have the same rank~\cite{Ox}. Since neutral and
charged cosmic strings have different ranks so they cannot be
related by a similarity transformation, consequently they do not
belong to the same class of equivalence.

Then b) by a perturbative analysis the search for asymptotic
solutions extending those derived in a). The perturbative analysis
consisted in retaining the first few terms of the power series of
$\chi$ in powers of $1/\rho$, that is, $\chi=(2/\rho)
A+2(\gamma/\rho^{2})D+2(\gamma/\rho^{3})E$ where the first term
$\chi_{0}=(2/\rho)A$ is the unique exact solution found in a) and
the 4$\times$4 constant real matrices $D$ \& $E$ had to be
determined solving the non-linear system (\ref{s} \& \ref{m}). We
ended by finding the exact non-trivial solution
\begin{eqnarray}
  \label{abd} \mbox{\hspace{-11mm}}& & \chi = (2/\rho)A-
  (4\gamma/\rho^{3})(A^{3}-A^{2})\,, \\
  \label{abd2} \mbox{\hspace{-11mm}}& &{\rm with}\;\aee =\be =\ee =1\,,
  \;\de=0\,,\;{\rm
  r}(A)=3\,,\nonumber
\end{eqnarray}
however, we failed to prove that it is mathematically the unique
polynomial solution subject to the constraints~\eqref{abd2}. The
perturbative approach has masked some features of the system that
we will recover here. In fact, the non-linear system (\ref{s} \&
\ref{m}) can be directly integrated leading to the most general
non-trivial solution (\ref{ns2}). Solution~\eqref{abd} has been
interpreted as an extended superconducting cosmic string
surrounding a longitudinally boosted electrically charged naked
cosmic string \cite{Azreg}.

In both solutions~\eqref{a} \& \eqref{abd} the invariants of $A$
are subject to the constraints
\begin{equation}\label{I}
    \aee =\be =\ee =1\,,\;\;\de=0\,,
\end{equation}
which reduce its characteristic equation to $A^{4}=A^{3}$. $A$ has
then the eigenvalues 1, 0, 0 \& 0 without necessarily being
diagonalizable. Instead, the simplest form to which $A$ can be
brought by a similarity transformation is the Jordan normal form
\cite{Ox}
\begin{equation}\label{J}
    A=\begin{pmatrix}
    1&0\\0&\mathbb{A}
\end{pmatrix}\,,\;\;
    \mathbb{A}=\begin{pmatrix}
    0&\epsilon_1&0\\0&0&\epsilon_2\\0&0&0
\end{pmatrix}
\end{equation}
where $\epsilon_1,\;\epsilon_2=0\;{\rm or}\;1$. Only the case
$\epsilon_1=0$ \& $\epsilon_2=~0$ corresponding to r$(A)=1$ is
diagonalizable. The case r$(A)=2$ corresponds either to
($\epsilon_1=0$ \& $\epsilon_2=1$) or to ($\epsilon_1=1$ \&
$\epsilon_2=0$) and the case r$(A)=3$ corresponds to
$\epsilon_1=\epsilon_2=1$. Now, with $A$ constrained by~\eqref{I},
it is straightforward to check that r$(A)=1$ or 2 implies
$A^{3}-A^{2}=0$, hence~\eqref{a} is just a special case
of~\eqref{abd} and the two solutions can be combined in one
\begin{equation}
\label{abd3} \chi = (2/\rho)A-(4\gamma/\rho^{3})(A^{3}-A^{2})
\end{equation}
(without constraining r$(A)$), where the invarinats of $A$ are
constrained by~\eqref{I}. The three cases with r$(A)=1$, 2 or 3 do
not belong to the same class of equivalence.

\section{Exact solutions}
The solution shown in (\ref{abd3}) is a special form of the more
general case of $\chi(\rho)$ being polynomial in a constant real
matrix $A$ with scalar coefficients. Using the characteristic
equation of $A$, one can eliminate any power of $A$ higher than 3,
hence the most general form of $\chi$ depending on a constant real
matrix is a polynomial in $A$ of third degree of the form
\begin{equation}\label{4}
    \chi(\rho)=\eta(\rho)+\omega(\rho)A+\beta(\rho)A^{2}+
    \delta(\rho)A^{3}\,.
\end{equation}
The constraints~\eqref{I} on the invariants of $A$ have ensured
the existence of neutral, charged and superconducting cosmic
strings so we will maintain them in our quest for new non-trivial
cosmic strings and focus on the determination of the four real
functions shown in (\ref{4}). Since the system (\ref{s} \&
\ref{m}) is non-linear, it may admit different solutions than
those shown in (\ref{abd3}).

\subsection{Case $\eta \equiv 0$}
First we assume that $\eta(\rho)\equiv0$ and re-parameterize
(\ref{4}) by $\beta(\rho)$, $\delta(\rho)$ and $T(\rho)\equiv{\rm
Tr}\chi(\rho)$
\begin{equation}\label{P4}
    \chi(\rho)=(T-\beta-\delta)(\rho)A+\beta(\rho)A^{2}+
    \delta(\rho)A^{3}\,.
\end{equation}

The set of conditions (\ref{I}) implies the relation $A^{n}=A^{3}$
for $n\geq3$ together with (\ref{P4}) reduce the system (\ref{s}
\& \ref{m}) to a scalar equation
\begin{equation}\label{p5}
    {\rm P}_{5}(\rho)\equiv 3[T_{,\rho}+(1/2)T^{2}]=0\,,
\end{equation}
along with a polynomial of degree 3 in $A$ with scalar
coefficients
\begin{equation}\label{p}
    {\rm P}_{4}(\rho)+{\rm P}_{3}(\rho)A+{\rm P}_{2}(\rho)A^{2}+
    {\rm P}_{1}(\rho)A^{3}=0\,,
\end{equation}
where
\begin{eqnarray}
  & &{\rm P}_{4}\equiv (2/3){\rm P}_{5}\label{p4} \\
  & &{\rm P}_{3}\equiv (1/3){\rm P}_{5}-(1/2)T(\beta+\delta)
  -(\beta+\delta)_{,\rho}\label{p3} \\
  & &{\rm P}_{1}+{\rm P}_{2}\equiv (1/2)T(\beta+\delta)+
  (\beta+\delta)_{,\rho}\label{p12}
\end{eqnarray}
\begin{eqnarray}
  {\rm P}_{1}\mbox{\hspace{-3mm}}&\equiv &\mbox{\hspace{-3mm}}
  (\gamma/4)T^{4}-(\gamma/2)
  (\beta+\delta)T^{3} \nonumber \\
  & &\mbox{\hspace{-3mm}} +\,[(3\gamma/2)T_{,\rho}-
  \gamma(\beta+\delta)_{,\rho}
  +(\gamma/4)(\beta+\delta)^{2}]T^{2} \nonumber \\
  & &\mbox{\hspace{-3mm}} +\,[(1/2)\delta-
  2\gamma(\beta+\delta)T_{,\rho}
    +\gamma(\beta+\delta)(\beta+\delta)_{,\rho}]T \nonumber \\
  & &\mbox{\hspace{-3mm}} +\,(\gamma/2)(\beta+\delta)^{2}T_{,\rho}+
  \delta_{,\rho}\,.
   \label{p1}
\end{eqnarray}
Eq. (\ref{p5}) being satisfied, the polynomial (\ref{p}) reduces
taking into account the relations (\ref{p4} to \ref{p12}) to
\begin{eqnarray}
   \mbox{\hspace{-5mm}}& &{\rm P}_{3}(\rho)A+{\rm P}_{2}(\rho)A^{2}+
    {\rm P}_{1}(\rho)A^{3}=0, \label{rp1} \\
   \mbox{\hspace{-5mm}}& &{\rm P}_{1}+{\rm P}_{2}=-{\rm P}_{3}=
   (1/2)T(\beta+\delta)+
  (\beta+\delta)_{,\rho}\,, \label{rp2}
\end{eqnarray}
and reduces further upon multiplying by $A$ and using (\ref{rp2})
along with $A^{n}=A^{3}$ for $n\geq3$ to
\begin{equation}\label{rp3}
    {\rm P}_{3}(\rho)(A^{2}-A^{3})=0\,.
\end{equation}
First, consider the case ${\rm P}_{3}\neq0$ \& $A^{3}=A^{2}$ which
leads upon substituting into (\ref{rp1}) to $A^{2}=A$. Now, the
relations $A^{3}=A^{2}=A$, leading to r$(A)=1$ or 2, reduce
(\ref{P4}) to $\chi=TA$ of the form (\ref{a}) already discussed.

We then expect new solutions from the other alternative
\begin{equation}\label{A1}
  {\rm P}_{3}=0=-(1/2)T(\beta+\delta)-
  (\beta+\delta)_{,\rho} \quad \& \quad A^{3}\neq A^{2}\,,
\end{equation}
with necessarly r$(A)=3$ ($\de=0$), leading taking into account
(\ref{rp2}) to
\begin{equation}\label{A2}
    {\rm P}_{2}=-{\rm P}_{1}\,.
\end{equation}
Now, equations (\ref{A1} \& \ref{A2}) together reduce (\ref{rp1})
to ${\rm P}_{1}~=~0=-{\rm P}_{2}$. Hence (\ref{p5}) along with the
alternative (\ref{A1}) lead to the vanishing of the coefficients
and the independent term of the polynomial (\ref{p}). We will now
proceed to the resolution of the equations ${\rm P}_{5}=0$, ${\rm
P}_{3}=0$ \& ${\rm P}_{1}=0$; the other equations, ${\rm P}_{4}=0$
\& ${\rm P}_{2}=0$, will be satisfied consequentially. For that
end we need to rewrite ${\rm P}_{1}=0$, where ${\rm P}_{1}$ given
by (\ref{p1}), in a simplified form using (\ref{p5}) to eliminate
$T_{,\rho}$ and (\ref{A1}) to eliminate $(\beta+\delta)_{,\rho}$.
Hence, when ${\rm P}_{5}=0$ \& ${\rm P}_{3}=0$ are satisfied we
have
\begin{equation}\label{p1s}
    {\rm P}_{1}=\delta_{,\rho}+(1/2)T\delta-
    (\gamma/2)T^{2}(T-\beta-\delta)^{2}=0\,.
\end{equation}

Equation (\ref{p5}) has two solutions $T=0$ \& $T=2/\rho$. If
$T=0$, (\ref{A1}) implies $\beta+\delta=c_{1}$ and (\ref{p1s})
implies $\delta=c_{2}$. Hence,
$\chi=-c_{1}A+(c_{1}-c_{2})A^{2}+c_{2}A^{3}$ is a constant matrix
case, which is already discussed in \cite{Azreg}.

Consider the other solution for $T$: $T=2/\rho$ where we have
omitted an insignificant additive constant in the denominator.
With this value of $T$, equation (\ref{A1}) too has two solutions
enumerated 1) \& 2) below.

1) A trivial solution to (\ref{A1}) is $\beta+\delta=0$ which
reduces (\ref{p1s}) to
\begin{equation}\label{p1ss}
    \delta_{,\rho}+(1/2)T\delta-
    (\gamma/2)T^{4}=0 \quad {\rm with} \quad T=2/\rho\,.
\end{equation}
Integrating this last equation, we obtain introducing a real
constant of integration $2K$
\begin{equation}
    \delta=-(4\gamma/\rho^{3})+2K/\rho \quad \& \quad
    \beta=-\delta,
\end{equation}
and then substituting into (\ref{P4}) we obtain
\begin{equation}\label{ns1}
    \chi=(2/\rho)A-(4\gamma/\rho^{3}-2K/\rho)(A^{3}-A^{2})\,.
\end{equation}
One sees that the solution (\ref{abd3}) is a special case of
(\ref{ns1}).

2) The other solution to (\ref{A1}) is proportional to $T$ given,
introducing a real constant of integration $L\neq0$, by
$\beta+\delta=2L/\rho=LT$. Substituting into (\ref{p1s}) and
introducing $\Gamma=~\gamma(1-L)^{2}$ we obtain the equation
$\delta_{,\rho}+~(1/2)T\delta-~(\Gamma/2)T^{4}=0$ (with
$T=2/\rho$) which is similar to (\ref{p1ss}) and hence has the
same solution
\begin{equation}
    \delta=-(4\Gamma/\rho^{3})+2K/\rho \quad \& \quad
    \beta=2L/\rho-\delta\,,
\end{equation}
Finally, substituting $T=2/\rho$ and the corresponding expressions
for $\delta$ \& $\beta$ shown in the previous equation into
(\ref{P4}) we obtain
\begin{eqnarray}
 \label{ns2} \chi &=& (2/\rho)[A+L(A^{2}-A)+K(A^{3}-A^{2})]\nonumber \\
   & & -\,[4\gamma(1-L)^{2}/\rho^{3}]
    (A^{3}-A^{2})\,.
\end{eqnarray}

Introducing the two relevant physical constants $I~=~L~-~K$ \&
$V=1-L$, the two solutions (\ref{ns1} \& \ref{ns2}) are combined
and rewritten as:

\subsection{Corollary} Assume $A$ a 4$\times$4 constant real
matrix of arbitrary rank satisfying the conditions (\ref{I}).
\begin{description}
    \item[] i) If $M=mA+nA^{2}+lA^{3}$ with
$m+n+l=1$, then $M$ satisfies (\ref{I}) and ${\rm r}(M)={\rm
r}(A)$.
    \item[] ii) If $M=VA+IA^{2}+(1-I-V)A^{3}$, $I$ \& $V$
$\in\mathbb{R}$, then
\begin{eqnarray}
  \label{ns3}\chi &=& (2/\rho)M-(4\gamma
V^{2}/\rho^{3})(A^{3}-A^{2}) \nonumber \\
   &=& (2/\rho)M-(4\gamma
/\rho^{3})(M^{3}-M^{2})
\end{eqnarray}
solves the system\footnote{The cases $L=0\;(V=1)$ \& $L\neq0
\;(V\neq1)$ correspond to~\eqref{ns1} \& \eqref{ns2},
respectively.} (\ref{s} \& \ref{m}). Since the rank of $M$ (or
$A$) is unconstrained, the different solutions are again
classified according to its values leading to three classes of
equivalence: neutral, charged and superconducting cosmic strings
for r$(M)=1$, r$(M)=2$ \& r$(M)=3$, respectively.
\end{description}

Now starting from the non-trivial solution (\ref{abd3}): $\chi=~
(2/\rho)A-(4\gamma/\rho^{3})(A^{3}-A^{2})$ with $A$ satisfying
(\ref{I}) \& r$(A)=3$ and performing a coordinate transformation
to the new coordinates $x_{\rm N}$ (\ref{L}), the thus
diffeomorphically obtained solution remains form-invariant:
$\chi_{\rm N}=~(2/\rho)A_{\rm N}-~(4\gamma/\rho^{3})(A_{\rm
N}^{3}-A_{\rm N}^{2})$ where $A_{\rm N}~=~S^{-1}AS$, $\chi_{\rm
N}=S^{-1}\chi S$. Since this latter solution $\chi_{\rm N}$ is of
the form (\ref{ns3}), one may try to solve for $S$ such that
$M\equiv~A_{\rm N}=~S^{-1}AS$. Taking
$M=~VA+~IA^{2}+~(1-~I-~V)~A^{3}$ and $A$ of the form (\ref{J})
with $\epsilon_1=\epsilon_2=1$ one obtains $S^2{}_2=S^3{}_3=1$,
$S^4{}_4=V$, $S^5{}_5=V^{2}$, $S^4{}_5=I$ and the other elements
$S^a{}_b$ vanish. Hence, with (\ref{I}) satisfied and
$\eta(\rho)\equiv 0$, the system (\ref{s} \& \ref{m}) admits no
further solutions of the form (\ref{4}) except those obtained by
coordinate transformations (\ref{L}) from~(\ref{abd3}). Although
the two solutions~\eqref{abd3} \& \eqref{ns3} are mathematically
diffeomorphic, however, they are only locally equivalent. The
similarity transformation $S$ derived above has the components
$S^5{}_5\neq \delta^5{}_5 =1$ \& $S^4{}_5\neq \delta^4{}_5 =0$,
this results in re-scaling the Killing vector ${\xi}_5{}^A$, which
has closed orbits, and mixing it with ${\xi}_4{}^A$. Hence, the
new solution~\eqref{ns3} is not globally equivalent
to~\eqref{abd3} \cite{Azreg}.

\subsection{Case $\eta \neq 0$}
Let us now assume that $\eta(\rho)\neq 0$. We will show that the
system (\ref{s} \& \ref{m}) does not admit any solution of the
form \eqref{4} if the constraints \eqref{I} are satisfied. Since
the latter lead to $A^{n}=A^{3}$ for $n\geq3$ we have then ${\rm
Tr}A^m=1$ for $m\geq1$ and consequently $A^m\neq0$ for $m\geq1$.
Now, substituting (\ref{I} \& \ref{4}) into the system (\ref{s} \&
\ref{m}) and using $A^{n}=A^{3}$ for $n\geq3$, the latter reduces
to
\begin{eqnarray}
   & & \mathbb{P}_{5}(\rho)=0\,;\label{A5} \\
   & &
   \mathbb{P}_{4}(\rho)+\mathbb{P}_{3}(\rho)A+\mathbb{P}_{2}(\rho)A^2+
   \mathbb{P}_{1}(\rho)A^3=0\,,\label{A6}
\end{eqnarray}
where $\mathbb{P}_{1}$ to $\mathbb{P}_{5}$ are expressed in terms
of the unknown functions $\eta$, $\omega$, $\beta$ \& $\delta$.
Multiplying (\ref{A6}) by $A^3$ and using once more $A^{n}=A^{3}$
for $n\geq3$ along with $A^m\neq0$ for $m\geq1$, we obtain
\begin{equation}\label{A7}
    \mathbb{P}_{4}+(\mathbb{P}_{3}+\mathbb{P}_{2}+
   \mathbb{P}_{1})=0\,.
\end{equation}
The trace of (\ref{A6}) leads to
\begin{equation}\label{A8}
 4\mathbb{P}_{4}+(\mathbb{P}_{3}+\mathbb{P}_{2}+
   \mathbb{P}_{1})=0\,.
\end{equation}
The homogeneous system (\ref{A7} \& \ref{A8}) in the two variables
$\mathbb{P}_{4}$ \&
($\mathbb{P}_{3}+\mathbb{P}_{2}+\mathbb{P}_{1}$) has only trivial
solutions
\begin{equation}\label{A9}
    \mathbb{P}_{4}=0\quad \& \quad (\mathbb{P}_{3}+\mathbb{P}_{2}+
   \mathbb{P}_{1})=0\,.
\end{equation}
Now, it is straightforward to show that the system of three
equations (\ref{A5} \& \ref{A9}) does not admit any solution if
$\eta(\rho)~\neq~0$. In (\ref{4}), re-parameterizing $\chi$ by
$\eta$, $\sigma=\omega+\beta+\delta$, $\beta$ \& $\delta$, the
system of three equations (\ref{A5} \& \ref{A9}) writes
\begin{eqnarray}
   & &\mathbb{P}_{5}\equiv [12\eta +3\sigma +(3/2)\gamma\sigma\eta^2+
   2\gamma\eta^3]_{,\rho} \nonumber \\
   \label{A11}& & +\,12\eta^2+6\eta\sigma +(3/2)\sigma^2 \\
   & & +3\,\gamma\eta^2[\eta^2+
   \eta\sigma +(1/4)\sigma^2]=0\nonumber \,;\\
   & &\mathbb{P}_{4}\equiv [9\eta +2\sigma +(5/2)\gamma\eta^3]_{,\rho} +
   \gamma[\eta^2\sigma_{,\rho} +7\sigma(\eta^2)_{,\rho}] \nonumber \\
   \label{A12}& & +\,12\eta^2+(11/2)\eta\sigma +\sigma^2\\
   & & +\,\gamma\eta^2[12\eta^2+
   (35/4)\eta\sigma +(5/4)\sigma^2]=0\nonumber \,;\\
   & &\mathbb{P}_{3}+\mathbb{P}_{2}+\mathbb{P}_{1}\equiv \sigma_{,\rho}
   +(1/2)\sigma^2+2\eta\sigma \nonumber \\
   \label{A13}& & +\,\gamma [3\eta(\sigma^2)_{,\rho}+(7/2)\eta^2\sigma_{,\rho} +
   (1/2)\sigma(\eta^2)_{,\rho}] \\
   & & +\,3\gamma\eta\sigma[13\eta^2+
   (43/4)\eta\sigma +(9/4)\sigma^2]=0\nonumber \,.
\end{eqnarray}
The three first order differential equations (\ref{A11} to
\ref{A13}), which do not depend on $\beta$ and $\delta$, have to
be solved for $\eta$ \& $\sigma$. When $\eta(\rho)\neq 0$, this
system will not admit any solution unless one of the three
equations is a combination of the two others. A way to check that
is to solve for the derivatives $\eta_{,\rho}$ \& $\sigma_{,\rho}$
using (\ref{A11}) \& (\ref{A13}) then substitute them into
(\ref{A12}) whose expression has to vanish identically. We have
checked it directly and found out that the expression of
$\mathbb{P}_{4}$ (\ref{A12}) reduces, but does not vanish
identically, to
\begin{eqnarray}
   2\gamma^2\mathbb{P}_{4} &=&
   \{3\gamma^2\eta(\gamma\eta^2+1)(2\eta+\sigma)^4 \nonumber \\
   & &+\,\gamma(5\gamma^2\eta^4+5\gamma\eta^2+2)(2\eta+\sigma)^3
   \nonumber \\
   & &
   +\,4\gamma\eta(5\gamma\eta^2-\gamma^2\eta^4+6)
   (2\eta+\sigma)^2 \\
   & &
   +\,(3\gamma^3\eta^6+6\gamma^2\eta^4+20\gamma\eta^2+8)
   (2\eta+\sigma)
   \nonumber \\
   & & -\,\gamma\eta^3(\gamma^2\eta^4+4\gamma\eta^2+4)\}/
   [\eta\sigma(2\eta+\sigma)]
   \,.\nonumber
\end{eqnarray}

\section{The superconducting cosmic string}
The 5-metric (\ref{metric}) is found upon integrating
$\chi\equiv{\lambda}^{-1}\,{\lambda}_{,\rho}$ using (\ref{ns2}) or
(\ref{ns3})
\begin{eqnarray}
  \lambda\mbox{\hspace{-3mm}} &=&\mbox{\hspace{-3mm}} C\{1-M^3+
  \rho^2 M^3-2\ln\rho (M^3-M) \nonumber \\
   \mbox{\hspace{-3mm}}&
   &\mbox{\hspace{-3mm}}-\,2[(\ln\rho)^2-\gamma/\rho^2](M^3-M^2)\}
   \nonumber \\
   \mbox{\hspace{-3mm}}&=&\mbox{\hspace{-3mm}} C\{1-A^3+\rho^2 A^3+
   2\ln\rho [VA+IA^{2}-(I+V)A^3] \nonumber \\
   \mbox{\hspace{-3mm}}&
   &\mbox{\hspace{-3mm}}-\,2V^{2}[(\ln\rho)^2-\gamma/\rho^2](A^3-A^2)\}\,.
   \label{l}
\end{eqnarray}
where $C$ is a constant real matrix of signature (-- -- + --).
Since $\lambda$ is symmetrical, $C$ satisfies the relations
$C=C^{\rm T}$ \& $CA=(CA)^{\rm T}$ (\textsuperscript{T} denotes
transpose). Choosing $A$ of the form~\eqref{J} with
$\epsilon_1=\epsilon_2=1$ (r$(A)=3$) \& $C$ of the form
\begin{equation}\label{333}
C=\begin{pmatrix} -\alpha^2&0\\0&\mathbb{C}
\end{pmatrix}\,,\;\;
\mathbb{C}=\begin{pmatrix} 0&0&-1\\0&-1&a\\-1&a&b
\end{pmatrix}
\end{equation}
the 5-metric (\ref{metric}) writes
\begin{eqnarray}
  {\rm d}s^{2}\mbox{\hspace{-3mm}}&=&\mbox{\hspace{-3mm}}-
  \,{\rm d}{\rho}^{2} - {\alpha}^{2}{\rho}^{2}\,{\rm d}{\varphi}^{2}
  -2\,{\rm d}z\,{\rm d}x^{5}\nonumber \\
\mbox{\hspace{-3mm}}& &\mbox{\hspace{-3mm}} -\,{\rm d}t^{2} -
4VQ(\rho)\,{\rm d}t\,{\rm d}x^5 \label{nm} \\
   \mbox{\hspace{-3mm}}& &\mbox{\hspace{-3mm}} -\,2[V^2Q(\rho)^2 +
   IQ(\rho) -\gamma V^2/\rho^2 - p]\,({\rm d}x^5)^2,\nonumber
\end{eqnarray}
where $p=(a^2+2b)/4-(aI/2V)$, $Q(\rho)=\ln(\rho/\rho_{0})$ and
$\rho_{0}$ is such that $a=2V\ln\rho_{0}$.

The metric~\eqref{nm} is subject to the following interpretation.
As we did earlier~\cite{Azreg}, reinterpreting the Lanczos term
$-(\gamma/8\pi G)L^{A}{_{B}}$ as a 5-dimensional effective
energy-momentum tensor source, equation~(\ref{field}) writes
$R^{A}{_{B}}-~(1/2)R\delta^{A}{_{B}}=8\pi GT^{A}{_{B}}$ where we
have inserted the gravitational constant. We evaluate the
components of $R^{A}{_{B}}$ as
\begin{eqnarray}
   \mbox{\hspace{-7mm}}& &R^{\mu}{_{\nu}}=(V\mathbb{A}+I\mathbb{A}^2)
   \Delta(\ln\rho)-\gamma
  V^2\mathbb{A}^2\Delta(1/\rho^2) \\
   \mbox{\hspace{-7mm}}& &R^{1}{_{1}}= R^{2}{_{2}}=[(\alpha -1)
  /\alpha]\Delta(\ln\rho)
\end{eqnarray}
where $\Delta$ is the covariant Laplacian with respect to the
metric ${\rm d}{\rho}^{2}+{\alpha}^{2}{\rho}^{2}\,{\rm
d}{\varphi}^{2}$ ($\mu$ \& $\nu$ assume the values from 3 to 5).
Using the relations $\Delta(1/\rho^2)=4/\rho^4$ \&
$\Delta(\ln\rho)=~2\pi\alpha\delta^2(\mathbf{x})$, where
$\mathbf{x}=\sqrt[\alpha]{\rho}\,(\cos\varphi , \sin\varphi)$ and
$\delta^2(\mathbf{x})$ is the 2-dimensional covariant
distribution, we obtain the non-vanishing components of the
5-dimensional effective energy-momentum tensor source by (since
${\rm Tr}\mathbb{A}=~{\rm Tr}\mathbb{A}^2=0$ then
$R=R^{A}{_{A}}=2R^{1}{_{1}}$):
\begin{eqnarray}
   & & T^{3}{_{3}}=T^{4}{_{4}}=T^{5}{_{5}}=\frac{(1-\alpha)}{4G}\,
   \delta^2(\mathbf{x}) \\
   & & T^{3}{_{4}}=T^{4}{_{5}}=\frac{\alpha V}{4G}\,
   \delta^2(\mathbf{x}) \\
   & & T^{3}{_{5}}=\frac{\alpha I}{4G}\,\delta^2({\mathbf{x}})-
   \frac{\gamma V^2}{2\pi
   G\rho^4}\,.
\end{eqnarray}
Hence, the tension ($T^{3}{_{3}}$), mass ($T^{4}{_{4}}$), scalar
charge ($T^{5}{_{5}}$), electric charge ($T^{4}{_{5}}$) and
momentum ($T^{3}{_{4}}$) densities of the string are purely
distributional. As is the case for local strings, there is no
tension in the radial direction ($T^{1}{_{1}}=0$) nor a pressure
in the $\varphi$-direction ($T^{2}{_{2}}=0$). The longitudinal
current density $T^{3}{_{5}}$ has a distributional contribution as
well as a continuous contribution proportional to $V^2$, which
diverges as $\rho$ approaches zero. For $V\neq0$, the 5-metric
(\ref{nm}) is then interpreted as an extended superconducting
cosmic string (continuous contribution of $T^{3}{_{5}}\to\infty$
as $\rho$$\to$0) surrounding a naked electrically charged cosmic
string core in longitudinal translation with a speed proportional
to $V$ and carrying an electric current proportional to $I$.

\section{Conclusion} Under cylindrical symmetry
assumption, the field equations of KK theory extended by GB term
reduce to the system (\ref{s} \& \ref{m}) of non-linear matrix and
scalar differential equations governing the behavior of a
4$\times$4 real matrix $\chi(\rho)$ whose elements are functions
of the 5-metric and its derivative. The system (\ref{s} \&
\ref{m}) may admit a variety of different solutions whose quest is
still an open question, in this letter we have shown that if
$\chi(\rho)$ is expressed as a polynomial in a constant real
matrix $A$ and if this is constrained by $\aee=\be=\ee=1$ \&
$\de=0$, then the system (\ref{s} \& \ref{m}) admits an exact
unique solution up to linear coordinate transformations mixing its
Killing vectors. Since the rank of $A$ is unconstrained, the
different solutions are classified according to its values in such
a way that solutions with the same rank of $A$ belong to the same
class of equivalence. Neutral and charged cosmic strings
correspond to r$(A)=1$ \& r$(A)=2$, respectively, while the
generic case r$(A)=3$ corresponds to superconducting cosmic
strings. 4-stationary solutions belonging to same class of
equivalence are not all globally equivalent due to the existence
of two Killing vectors with closed orbits. For r$(A)=3$, we have
shown that the most general superconducting cosmic string is in
arbitrary translational motion parallel to its axis, carries an
arbitrary electric current and is surrounded by a continuous
longitudinal current density.

The question whether the non-linear system (\ref{s} \& \ref{m})
admits other polynomial solutions in a constant matrix $A$,
however, constrained by other conditions than those shown above is
still an open topic. On the other hand, there is no reason to
restrict oneself to solutions depending on a constant matrix; are
there solutions of different shapes, for instance solutions
depending on two constant matrices? In a project that we are
realizing, these questions are only party answered: we managed to
prove that any solution $\chi(\rho)$ to the system (\ref{s} \&
\ref{m}) is necessarily a polynomial in a constant real matrix
with scalar coefficients depending on $\rho$~\cite{Azreg2}. This
restricts the search for new solutions to polynomials.

\end{document}